%% file: EXO-21-003_temp.tex
\pdfoutput=1
\documentclass[11pt,twoside,a4paper,cmspaper,final,collab]{cms-tdr}
\def\svnVersion{0481bee}\def\svnDate{2023/07/11}\def\cmsCernNoTag{CERN-EP-2022-105}\def\cmsCernDate{\today}\def\cmsMessage{Published in Physical Review Letters as \href{http://dx.doi.org/10.1103/PhysRevLett.131.011803}{\doi{10.1103/PhysRevLett.131.011803}.}}
\begin{document}\cmsNoteHeader{EXO-21-003}

\cmsNoteHeader{EXO-21-003}
\newcommand{\dphill}{\ensuremath{\Delta\phi_{\ell\ell}}\xspace}
\newcommand{\WW}{\ensuremath{W^{\pm} W^{\pm}}\xspace}
\newcommand{\WZ}{\ensuremath{W^{\pm} Z}\xspace}
\newcommand{\ZZ}{\ensuremath{Z Z}\xspace}
\newcommand{\TTBar}{\ensuremath{t\bar{t}}\xspace}

\newcommand{\hmn}{heavy Majorana neutrino\xspace}
\newcommand{\mN}{m_{N}\xspace}
\newcommand{\WO}{Weinberg operator\xspace}
\newcommand{\mjj}{m_{jj}}
\newcommand{\Vln}{V_{\ell N}\xspace}
\newcommand{\Vlnprime}{V_{\ell' N}\xspace}
\newcommand{\Vmn}{V_{\PGm N}\xspace}
\newcommand{\pT}{p_{T}\xspace}
\newcommand{\pTmiss}{p_{T}^{\mathrm{miss}}\xspace}
\newcommand{\pTvecmiss}{\ensuremath{{\vec p}_{T}^{\kern1pt\text{miss}}}\xspace}
\newcommand{\hT}{\ensuremath{H_{T}}\xspace}
\newcommand{\lag}{\ensuremath{\mathcal{L}}\xspace}

\title{Probing heavy Majorana neutrinos and the Weinberg operator through vector boson fusion processes in proton-proton collisions at \texorpdfstring{$\sqrt{s}~=~13~\TeV$}{sqrt(s) = 13 TeV}}

\date{\today}

\abstract{The first search exploiting the vector boson fusion process to probe heavy Majorana neutrinos and the Weinberg operator at the LHC is presented. The search is performed in the same-sign dimuon final state using a proton-proton collision dataset recorded at $\sqrt{s}=13\TeV$, collected with the CMS detector and corresponding to a total integrated luminosity of 138\fbinv. The results are found to agree with the predictions of the standard model. For heavy Majorana neutrinos, constraints on the squared mixing element between the muon and the heavy neutrino are derived in the heavy neutrino mass range $50\GeV$--$25\TeV$; for masses above 650\GeV these are the most stringent constraints from searches at the LHC to date. A first test of the Weinberg operator at colliders provides an observed upper limit at 95\% confidence level on the effective $\PGm\PGm$ Majorana neutrino mass of 10.8\GeV.}

\hypersetup{
pdfauthor={CMS Collaboration},
pdftitle={Probing Majorana neutrinos and Weinberg operator in di-muon channel through vector boson fusion processes in proton-proton collisions at 13 TeV},
pdfsubject={CMS},
pdfkeywords={CMS, Majorana neutrino, Weinberg operator, vector boson fusion}}

\maketitle
Despite the enormous success of the standard model (SM) of particle physics over the last fifty years, the observation of neutrino oscillations~\cite{SNO:2002tuh,Super-Kamiokande:2005mbp} has confirmed that at least two SM neutrinos have small, but nonzero, masses. This provides a compelling hint of physics beyond the SM. Efforts to determine each neutrino mass are being pursued by many experiments within different research fields, including cosmology and tritium beta decay~\cite{Loureiro:2018pdz,eBOSS:2020yzd,KATRIN:2021uub}, and various theoretical explanations of these observations have been proposed~\cite{seesaw1,seesaw2,seesaw3,seesaw4,Weinberg79,Das:2015toa,Das:2016hof,Das:2017nvm}.

A natural formalism for generating neutrino masses is through a dimension-five operator, proposed by Weinberg~\cite{Weinberg79}, which extends the SM Lagrangian with
\begin{linenomath}
\begin{equation}
    \lag_5 = \frac{\mathrm{C}_5^{\ell\ell'}}{\Lambda} \big[\Phi\!~\cdot~\! \overline{L}^c_{\ell }\big] \big[L_{\ell'}\!\!~\cdot~\!\Phi\big],
    \label{eqn:Lag_5}
\end{equation}
\end{linenomath}
where $\ell ,~\ell'$ are the flavors of the leptons, which can be electrons, muons, or taus; $\Lambda$ is the scale at which the particles responsible for neutrino masses become relevant degrees of freedom; $\mathrm{C}_5^{\ell\ell'}$ is a flavor-dependent Wilson coefficient; $L_\ell=(\nu_\ell,\ell)$ is the left-handed lepton doublet; and $\Phi$ is the SM Higgs doublet with a vacuum expectation value $v=\sqrt{2}\langle\Phi\rangle\approx246\GeV$. This operator, known as the Weinberg operator, generates the Majorana neutrino masses as
\begin{linenomath}
\begin{equation}
    m_{\ell\ell'} = C_5^{\ell\ell'} v^2/\Lambda,
    \label{eqn:Majorana_N_mass}
\end{equation}
\end{linenomath}
and introduces lepton number violation (LNV), as the Majorana neutrino is its own antiparticle. The LNV processes have been tested extensively in searches for neutrinoless double beta decays of heavy nuclei. These experiments search for events with two same-sign (SS) charge electrons and the absence of neutrinos in the final state, and set stringent upper limits on the associated Majorana neutrino mass ($ \abs{m_{ee}}$) at 90\% confidence level (\CL) in the ranges 61--165\unit{meV}~\cite{KamLAND-Zen:2016pfg} and 79--180\unit{meV}~\cite{Agostini:2020xta}.
Since muons and tau leptons are much heavier than electrons, SS final states involving these heavier leptons are kinematically forbidden in low-energy nuclear experiments and can only be studied at colliders~\cite{Fuks:2020zbm}. As the study of tau leptons is experimentally more challenging, the final state with SS muons is investigated in this search. The observed upper limit on the effective $\PGm\PGm$ Majorana neutrino mass ($\abs{m_{\PGm\PGm}}$), translated from the searches for LNV decays of charged kaons at the NA62 experiment~\cite{NA62:2019eax}, is around 55\GeV at 90\% \CL~\cite{Fuks:2020zbm}.

The Weinberg operator can be realized in the context of ``seesaw'' models~\cite{seesaw1,seesaw2,seesaw3,seesaw4}, assuming the existence of hypothetical heavy states, for example a heavy Majorana neutrino ($\mathit{N}$) in the type-I seesaw model. The small mass of the SM neutrinos can thus be explained by a suppression due to the high mass of new particles, because $ m_\nu\sim y_\nu^2v^2/ \mN$. In this expression, $m_\nu$ is the SM neutrino mass, $ \mN$ is the mass of the heavy Majorana neutrino, and $y_\nu$ is the Yukawa coupling. This ability to explain the small mass of the SM neutrinos motivates the search for heavy Majorana neutrinos. Heavy Majorana neutrinos can only couple to the SM through mixing with SM neutrinos, which is characterized by the mixing element, $\Vln$, between a SM neutrino in its left-handed interaction state and a \hmn in its mass eigenstate. The searches that have set limits on $\Vln$ were based on the production of a \hmn through either quark-antiquark annihilation or $W\PGg$, with different analysis strategies based on various decay modes, such as many leptons in the final state~\cite{Sirunyan:2018mtv,Aad:2019kiz,Sirunyan:2019bgz}, and in final states containing LNV lepton pairs~\cite{Sirunyan:2018xiv,Aaboud:2018spl}. For $ \mN$ at a few \GeVns{}, the constraints on $\abs{\Vln}^{2}$, where $\ell=\left\{e, \Pgm \right\}$, are around $10^{-7}$--$10^{-5}$, while at around 1\TeV, the constraints approach unity.

\begin{figure}[h!]\centering
\includegraphics[width=0.40\linewidth]{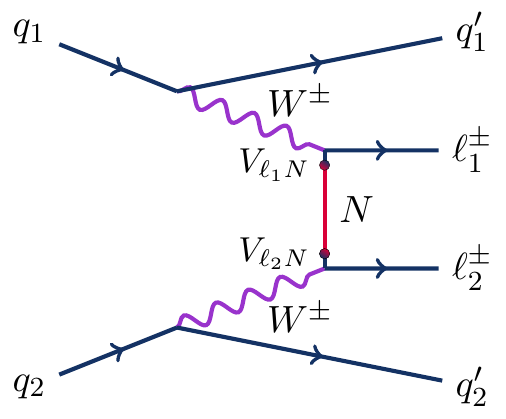}
\hspace{13pt}
\includegraphics[width=0.40\linewidth]{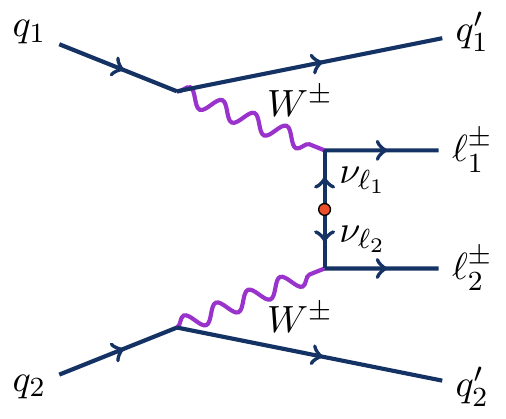}
\caption{Example Feynman diagrams of VBF processes with heavy Majorana neutrino production (left) and processes mediated by the Weinberg operator (right) at the LHC.}\label{fig:diagram}
\end{figure}

This Letter reports the first search probing Majorana neutrinos and the Weinberg operator in vector boson fusion (VBF) processes~\cite{Fuks:2020zbm,Fuks:2020att} at the CERN LHC, studying events with SS muon pairs.
As shown in Fig.~\ref{fig:diagram} by the two representative Feynman diagrams, in the $t$ channel, SS $W$ boson pairs may convert to SS lepton pairs via a \TeVns{}-scale Majorana neutrino or through a Weinberg operator process. Because the cross section of $t$-channel processes is less sensitive to the mass of the intermediate particle compared with $s$-channel quark-antiquark annihilation processes, the VBF process can complement searches for heavy Majorana neutrinos at the \TeVns{} mass scale as its cross section decreases more slowly with increasing $N$ mass compared with the values from $s$-channel production. Although the various seesaw mechanisms can be considered as realizations of the Weinberg operator, an alternative implementation is considered in this Letter, as shown in Fig.~\ref{fig:diagram} (right). The two VBF processes considered in this work differ primarily in that the Majorana process is assumed to be mediated by a heavy $t$-channel neutrino whereas the Weinberg operator process is mediated by a lighter $t$-channel neutrino.

This analysis is based on proton-proton ($pp$) collision data collected at $\sqrt{s} = 13\TeV$ by the CMS experiment at the CERN LHC during 2016--2018, corresponding to an integrated luminosity of $138\fbinv$.

Tabulated results are provided in the HEPData record for this analysis~\cite{hepdata}.

The CMS apparatus~\cite{CMS:2008xjf} is a multipurpose, nearly hermetic detector, designed to trigger on~\cite{CMS:2020cmk,CMS:2016ngn}
and identify electrons, muons, photons, and hadrons~\cite{CMS:2020uim,CMS:2018rym,CMS:2014pgm}. A global ``particle-flow'' algorithm~\cite{CMS:2017yfk} aims to reconstruct all individual particles in an event, combining information provided by the all-silicon inner tracker and by the crystal electromagnetic and brass-scintillator hadron calorimeters, operating inside a 3.8\unit{T} superconducting solenoid, with data from gas-ionization muon detectors embedded in the flux-return yoke outside the solenoid. The reconstructed particles are used to build charged leptons, jets, and missing transverse momentum ($\pTvecmiss$)~\cite{CMS:2018jrd,CMS:2016lmd,CMS:2019ctu}.

Samples of signal events are simulated with next-to-leading order (NLO) precision using the \MGvATNLO 2.6.5 generator~\cite{Alwall:2014hca}, based on model implementations from Refs.~\cite{Fuks:2020zbm,Alva:2014gxa,Degrande:2016aje,Fuks:2020att}. For the heavy Majorana neutrino, we assume only one heavy state and scan its mass in a range from 50\GeV to 25\TeV. The process induced by the Weinberg operator is simulated using a novel approach, established in Ref.~\cite{Fuks:2020zbm}, which approximates the internal neutrino lines and vertex insertion with an effective light Majorana neutrino. The Wilson coefficient $\mathrm{C}_5^{\ell\ell'}$ is set to 1, and the energy scale $\Lambda$ to 200\TeV. The production cross sections scale with $\Vln ^{2} \Vlnprime^{2}$ and $\abs{\mathrm{C}_5^{\ell\ell'}/\Lambda }^{2}$ for the VBF processes with heavy Majorana neutrinos~\cite{Fuks:2020att} and the Weinberg operator~\cite{Fuks:2020zbm}, respectively. Since the kinematic shape of the signal is independent of $\abs{ \Vmn }$, we set $\abs{ \Vmn } = 1$ for the signal simulation to ensure efficient generation of signal events.

The SM electroweak $\WW$ and $W^{\pm} Z$ processes, where both bosons decay leptonically, are simulated using \MGvATNLO 2.4.2~\cite{Alwall:2014hca} at leading order (LO) with six electroweak $\left[ \mathcal{O}(\alpha^6) \right]$ and zero quantum chromodynamics (QCD) vertices. The QCD-induced $\WW$ process is also simulated using \MGvATNLO 2.4.2 at LO. Contributions with an initial-state $b$ quark are excluded from the electroweak $W^{\pm} Z$ simulation because they are included in the production of a $Z$ boson in association with a single top quark, known as the $tZq$ process, which is simulated at NLO using \MGvATNLO 2.3.3.
The $ZZ \to 4 \ell$ and $gg \to ZZ$ processes are simulated with the $\POWHEG$ v2~\cite{Nason:2013ydw} and $\MCFM$ 7.0.1~\cite{Campbell:2011bn} generators, respectively. The production of $\TTBar W^{\pm}$, $\TTBar Z$, and triple vector boson ($VVV$) background events is simulated at NLO in QCD using the \MGvATNLO 2.2.2 (2.4.2) generator~\cite{Alwall:2014hca,MadGraph,Frederix:2012ps} for the samples corresponding to the 2016 (2017--2018) data-taking period. The production of $\WW$ events through double-parton scattering is generated at LO using \PYTHIA 8.226 (8.230)~\cite{Sjostrand:2014zea} for 2016 (2017--2018).

The simulated samples for the 2016 (2017--2018) data-taking period use the NNPDF 3.0 (3.1) parton distribution functions~\cite{Ball:2014uwa,Ball:2017nwa}, and are interfaced with \PYTHIA to model the fragmentation and hadronization of partons in the initial and final states, along with the underlying event. The CUETP8M1 (CP5) tune~\cite{Khachatryan:2015pea,Sirunyan:2019dfx} is used in the simulation for the 2016 (2017--2018) data-taking period. In LO (NLO) simulations performed with \MGvATNLO, jets are matched to the parton shower produced by \PYTHIA 8.2 following the MLM~\cite{Alwall:2007fs} (FxFx~\cite{Frederix:2012ps}) prescription. The interactions of all final-state particles with the CMS detector are simulated using \GEANTfour~\cite{Agostinelli:2002hh}. Simulated events are mixed with the contribution of particles from additional $pp$ interactions within the same or nearby bunch crossings (pileup); the pileup distribution in the simulated samples matches that observed in data.

Collision events are collected using single-muon triggers~\cite{CMS:2016ngn} that require the presence of an isolated muon with transverse momentum $\pT > 24$ or 27\GeV, depending on the data-taking period. In addition, a set of dimuon triggers~\cite{CMS:2016ngn} with lower $\pT$ thresholds is used to ensure a trigger efficiency above 90\% for events with muons. Depending on the subsequent off-line selection, the dimuon triggers increase the event acceptance by up to 4\%. The primary vertex (PV) is taken to be the vertex corresponding to the hardest scattering in the event, evaluated using tracking information alone, as described in Section 9.4.1 of Ref.~\cite{CMS-TDR-15-02}.
The jets are clustered using the anti-\kt algorithm~\cite{Cacciari:2011ma,Cacciari:2008gp} with a distance parameter of 0.4, with the tracks assigned to the candidate vertices as inputs. To mitigate the impact from pileup, tracks identified to be originating from pileup vertices are discarded and calorimeter energies are corrected to account for charged and neutral particles originating from those vertices. The $\pTvecmiss$ is then defined as the negative vector $\pT$ sum of all reconstructed particles in an event. Its magnitude is referred to as $\pTmiss$. Jets arising from $b$ quark hadronization and decay ($b$ jets) are identified using a deep neural network algorithm, termed \textsc{DeepCSV}, whose inputs are tracks displaced from the primary interaction vertex, details of secondary vertices, and the kinematical variables of jets~\cite{Sirunyan:2017ezt}. To classify jets as $b$ tagged, we use a requirement on the score of the \textsc{DeepCSV} algorithm for which the $b$ jet identification efficiency ranges from 65 to 75\% and the rate at which gluon or light-flavor jets are misidentified as $b$ jets is around 1\%. The fraction of jets originating from the hadronization of $c$ quarks that are incorrectly identified as $b$ jets is about 10\%.
Starting from the constituents of reconstructed jets, hadronically decaying $\tau$ leptons are identified via the ``hadrons-plus-strips'' algorithm~\cite{CMS:2018jrd}.

The final states analyzed contain two well-identified SS muons with $\pT > 30\GeV$ and pseudorapidity $\abs{\eta} < 2.4$, and at least two jets with $\pT > 30\GeV$ and $\abs{\eta} < 4.7$. The two selected jets with the largest $\pT$ are considered as the VBF jet candidates, which must satisfy a large pseudorapidity separation $\abs{\Delta\eta_{jj}} > 2.5$ and a large dijet invariant mass $m_{jj} > 750 \GeV$. Events produced via VBF are expected to contain rapidity gaps between the muons and the VBF jets. To utilize this signature, we require max$\left ( \mathcal{Z}_{\ell} \right ) < 0.75$, where $\mathcal{Z}_{\ell}$ is the Zeppenfeld variable~\cite{zepp1996} of a lepton $\ell$: $\mathcal{Z}_{\ell} = \abs{ \eta_{\ell} -(\eta_{j1}+\eta_{j2})/2 }/  \abs{ \Delta \eta_{jj} }$, where $\eta_{\ell}$ is the pseudorapidity of the lepton, and $\eta_{j1}$ and $\eta_{j2}$ are of the two VBF jets. Events with an additional loosely identified muon (electron) with $\pT > 10 \GeV$ and $\abs{\eta} < 2.5 (2.4)$ or an identified hadronically decaying tau lepton with $\pT > 20 \GeV$ and $\abs{\eta}<2.3$ are rejected to suppress backgrounds such as $W^{\pm} Z$ production. Because the signal processes do not produce final state neutrinos other than those produced in hadron decays, $\pTmiss$ is a useful discriminant to separate the signal and backgrounds. For the analysis targeting processes with heavy Majorana neutrinos, the azimuthal separation variable $\dphill$ is better for improving the sensitivity to the signal than $\pTmiss$ because the two muons in the final state tend to be more back-to-back than in background processes~\cite{Fuks:2020att}. However, $\pTmiss$ is more discriminating than azimuthal separation for the Weinberg operator process because of the different event topology resulting from the discrepancy between the two $t$-channel propagators. To separate the signal and the SM electroweak $\WW$ events, signal regions (SRs) for the \hmn and Weinberg operator analyses are defined in bins of $\dphill$ and $\pTmiss$, respectively. For the \hmn analysis, we define a high-$\dphill$ bin ($\dphill > 2$) and a low-$\dphill$ bin ($\dphill < 2$) in the SR. For the Weinberg operator analysis, a low-$\pTmiss$ bin ($<~$50\GeV) and a high-$\pTmiss$ bin ($>~$50\GeV) are defined in the SR.

The background from opposite-sign dilepton events where one of the lepton charges is misidentified is negligible for events with muon pairs~\cite{CMS:2009fdy,CMS:2019ied}. The so-called nonprompt-lepton backgrounds, originating from leptonic decays of heavy quarks or hadrons misidentified as leptons, are suppressed by identification and isolation requirements.
Events from a $\TTBar$ pair where only one $W$ boson decays leptonically are the main source of nonprompt-lepton backgrounds. To reduce this contribution, events with at least one $b$-tagged jet with $\pT > 20\GeV$ and $\abs{\eta}<2.4$ are rejected. The remaining contribution of the nonprompt-lepton background is estimated directly from a data sample by applying weights to events containing muon candidates that fail the nominal selection criteria while passing a less stringent isolation requirement. These weights, called fake-lepton factors, are obtained from the probability of a jet being misidentified as a lepton and the efficiency of correctly reconstructing and identifying a lepton. More details about this method are given in Refs.~\cite{Sirunyan:2018egh,CMS:2020dvg}.

To constrain the reducible background contributions, which include $W^{\pm} Z$ production and backgrounds from nonprompt leptons, several control regions (CRs) in data are used. These CRs are defined to select samples of events enriched in those background processes, following the procedure described in Refs.~\cite{Sirunyan:2017ret,Sirunyan:2020gyx,Sirunyan:2020gvn}.
A $WZ$ CR, requiring the presence of three muons in an event, is used to estimate $W^{\pm} Z$ background contributions. The opposite-sign dimuon combination with invariant mass closest to the $Z$ boson mass ($m_{Z}$) is considered as the $Z$ boson decay product, whereas the remaining muon is assumed to originate from the decay of the $W$ boson. The selection criteria defining the SR and the $WZ$ CR are summarized in Table~\ref{tab:regions}.
To select event samples enriched in nonprompt leptons, a $b$-tagged CR is defined by using the same selection as for the SR, however requiring the presence of a $b$-tagged jet. Similarly, the $WZb$ CR is defined by requiring the same selection as for the WZ CR, but requiring at least one $b$-tagged jet. The dominant backgrounds in the SR are SM electroweak $\WW$ production and the contribution from nonprompt leptons.

\begin{table*}[htbp]
    \centering
\topcaption{Signal region and $WZ$ control region definitions. The signal region is binned differently for the heavy Majorana neutrino and Weinberg operator analyses. A $b$-tagged CR is defined by using the same selection as for the SR, however requiring the presence of a $b$-tagged jet. A $WZb$ CR is defined by requiring the same selection as for the $WZ$ CR, but requiring at least one $b$-tagged jet.}
\begin{scotch}{l c c}
Variable & SR & $WZ$ CR \\
\hline
Number of leptons                          & Exactly 2 same-sign muons & Exactly 3 muons \\
Lepton $\pT$                                 & $>~$30/30~\GeV            & $>~$25/25/10~\GeV \\
Lepton \abs{\eta}                          & $<~$2.4                  & $<~$2.4 \\
Jet $\pT$                                    & $>~$30~\GeV               & $>~$30~\GeV \\
Jet \abs{\eta}                             & $<~$4.7                  & $<~$4.7 \\
Dilepton mass $m_{\ell \ell}$              & $>~$20~\GeV               & \NA  \\
$\abs{m_{\ell \ell}-m_{Z}}$              & \NA                     & $<~$15~\GeV (for one $\PGmp\PGmm$ pair) \\
Trilepton mass $m_{\ell \ell \ell}$        & \NA                     & $>~$100~\GeV \\
Max $\left ( \mathcal{Z}_{\ell} \right )$  & $<~$0.75                 & $<~$1.0      \\
$\pTmiss$                                    & \NA                     & $>~$30~\GeV \\
$b$-tagged jet veto                       & Required                & Required \\
Hadronic $\tau$ veto                       & Required                & Required \\
$\mjj$                                     & $>~$750~\GeV              & $>~$750~\GeV \\
$\abs{\Delta\eta_{jj}}$                    & $>~$2.5                  & $>~$2.5 \\
\end{scotch}
    \label{tab:regions}
\end{table*}

The discriminating variable used to search for a signal for both analyses is $\hT/\pT^{\PGm_1}$. Here, $\hT$ is the scalar sum of jet $\pT$ for all jets with $\pT>30\GeV$ and $\abs{\eta} < 4.7$, $\PGm_1$ indicates the highest $\pT$ muon.
This discriminating variable measures the amount of hadronic activity relative to leptonic activity in a given event, and is sensitive to the color structure of the hard scattering processes as discussed in Ref.~\cite{Fuks:2020zbm}. As hadronic activity is typically suppressed in VBF production processes, $\hT/\pT^{\PGm_1}$ tends to be smaller for signal events than for background events~\cite{Fuks:2020zbm,Fuks:2020att}.

The statistical analysis is performed with a profile likelihood ratio test statistic in which systematic uncertainties are modeled as constrained nuisance parameters. Limits are set using the modified frequentist \CLs criterion~\cite{CLS2,CLS1}, and the asymptotic approximation~\cite{Cowan:2010js} is used in the limit setting procedure. The integrated luminosities of the 2016, 2017, and 2018 data-taking periods are individually known with uncertainties in the range 1.2--2.5\%~\cite{CMS:2021xjt,CMS:2018elu,CMS:2019jhq}, while the uncertainty in the total integrated luminosity in 2016--2018 is 1.6\%. The simulation of pileup events assumes a total inelastic $pp$ cross section of 69.2\unit{mb}, with an associated uncertainty of 5\%~\cite{ATLAS:2016pu,Sirunyan:2018nqx}, which has an impact of approximately 1\% on the expected signal and background yields. Discrepancies in the lepton reconstruction and identification efficiencies between data and simulation are corrected for by applying scale factors to all simulated samples, and the associated uncertainty is around 2\%. Uncertainties in both the muon momentum scale and resolution individually amount to about 0.2\%. The simulated muon momentum resolution is corrected by applying a 15\% Gaussian smearing to muons with $\pT > 200\GeV$~\cite{CMS:2019ied}.

The resolution uncertainty and the momentum scale uncertainty for high-$\pT$ muons are less than 1\%. The uncertainty in the calibration of the jet energy scale (JES) directly affects the acceptance of the jet multiplicity requirement and the $\pTmiss$ measurement. These effects are estimated by shifting the JES in the simulation up and down by 1 standard deviation. The uncertainty in the JES is 2\%--5\%, depending on the jet $\pT$ and $\eta$~\cite{CMS:2016lmd,CMS-DP-2020-019}. Uncertainties in the $b$ tagging algorithm are introduced by various sources~\cite{Sirunyan:2017ezt}, and the overall impact on the simulated samples is less than 1\%. Theoretical uncertainties arising from the choice of the renormalization and factorization scales are estimated by varying these scales independently up and down by a factor of 2 from their nominal values, excluding the two extreme variations~\cite{Biedermann:2017bss}. The largest cross section variations from this procedure are taken as the uncertainty~\cite{Ballestrero:2018anz}, which is about 6\%, and which is applied to both the signal and the SM electroweak $\WW$ processes. The parton distribution function uncertainty of the signal and the SM electroweak $\WW$ processes is evaluated according to the procedure described in Ref.~\cite{PDFLHC}, and is found to be about 1\%. These uncertainties are not applied to the subdominant $\WZ$ background, where they are not significant compared to other uncertainties considered. For the nonprompt-lepton background prediction, systematic uncertainties arise from the limited size of the sample used to measure the efficiency and from the difference in the flavor composition of the jets that are misidentified as leptons between the measured sample and the signal region, estimated to amount to at most 5 and 7\%, respectively. An additional normalization uncertainty of 30\% is assigned to the nonprompt-lepton background, based on studies of simulated events~\cite{CMS:2020dvg}.
The statistical uncertainty in the signal and background templates is introduced as a systematic uncertainty via the Barlow-Beeston-lite approach~\cite{Barlow:1993dm}. This is the systematic uncertainty that has the largest impact on the results derived from the simultaneous fits described below. Uncertainties related to pileup, theoretical calculations, and the muon trigger and identification are correlated across the three data-taking years. The luminosity and $b$ tagging uncertainties are partially correlated through some of the individual uncertainty sources, and the remaining uncertainties are uncorrelated across the three datasets. The total uncertainties on the results are dominated by the statistical uncertainties.

\begin{figure*}[htb!]
    \centering
    \includegraphics[width=1\textwidth]{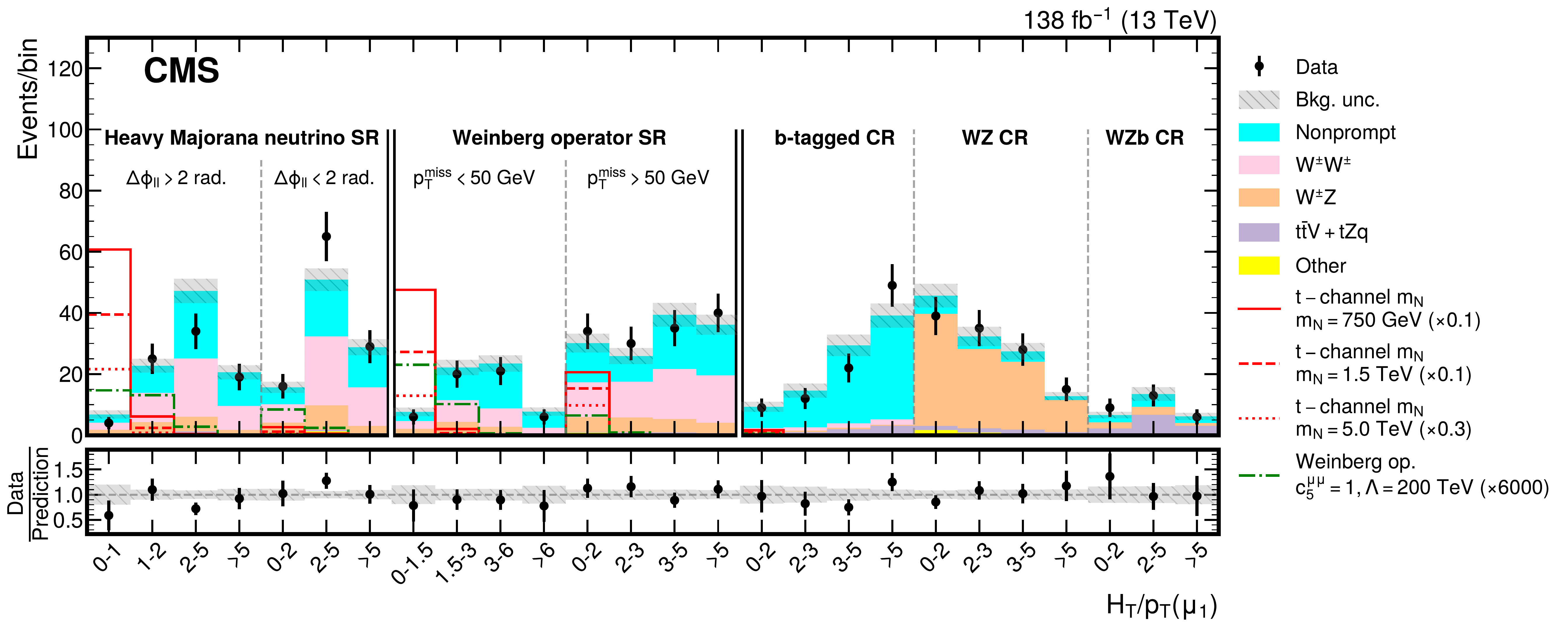}
    \caption{The $\hT/\pT^{\PGm_{1}}$ distribution in the SRs and CRs. The horizontal axis indicates the bin width. The predicted yields of the backgrounds are shown with their best fit normalizations from the simultaneous fits for the background-only hypothesis. The lines indicate the scaled expected distributions of the \hmn process with $\mN = 750\GeV$ (solid red line), 1.5\TeV (dashed red line), 5.0\TeV (dotted red line), and of the \WO process (dash-dotted green line). The \hmn processes are characterized by the azimuthal separation variable $\dphill$, while the \WO process features in low missing transverse momentum.}
    \label{fig:signal_region}
\end{figure*}

Two separate fits are performed, based on the SR and the $b$-tagged, $WZ$, and $WZb$ CRs: one for the heavy Majorana neutrino analysis using $\dphill$ bins in the SR, and a second for the Weinberg operator analysis with the $\pTmiss$ bins in the SR. The normalization factors for the $\WW$, $\WZ$, and $tZq$ background processes, affecting both the SRs and CRs, are included as free parameters in the fit together with the signal strength. The predicted yields are shown in Fig.~\ref{fig:signal_region} with their best fit normalizations from the simultaneous fits for the background-only hypothesis, \ie assuming no contributions from processes involving heavy Majorana neutrinos and the Weinberg operator. The bin boundaries are chosen to optimize the signal sensitivity while leaving the major background contributions in each bin at a reasonable level.

\begin{figure}[htbp!]
    \centering
    \includegraphics[width=0.50\textwidth]{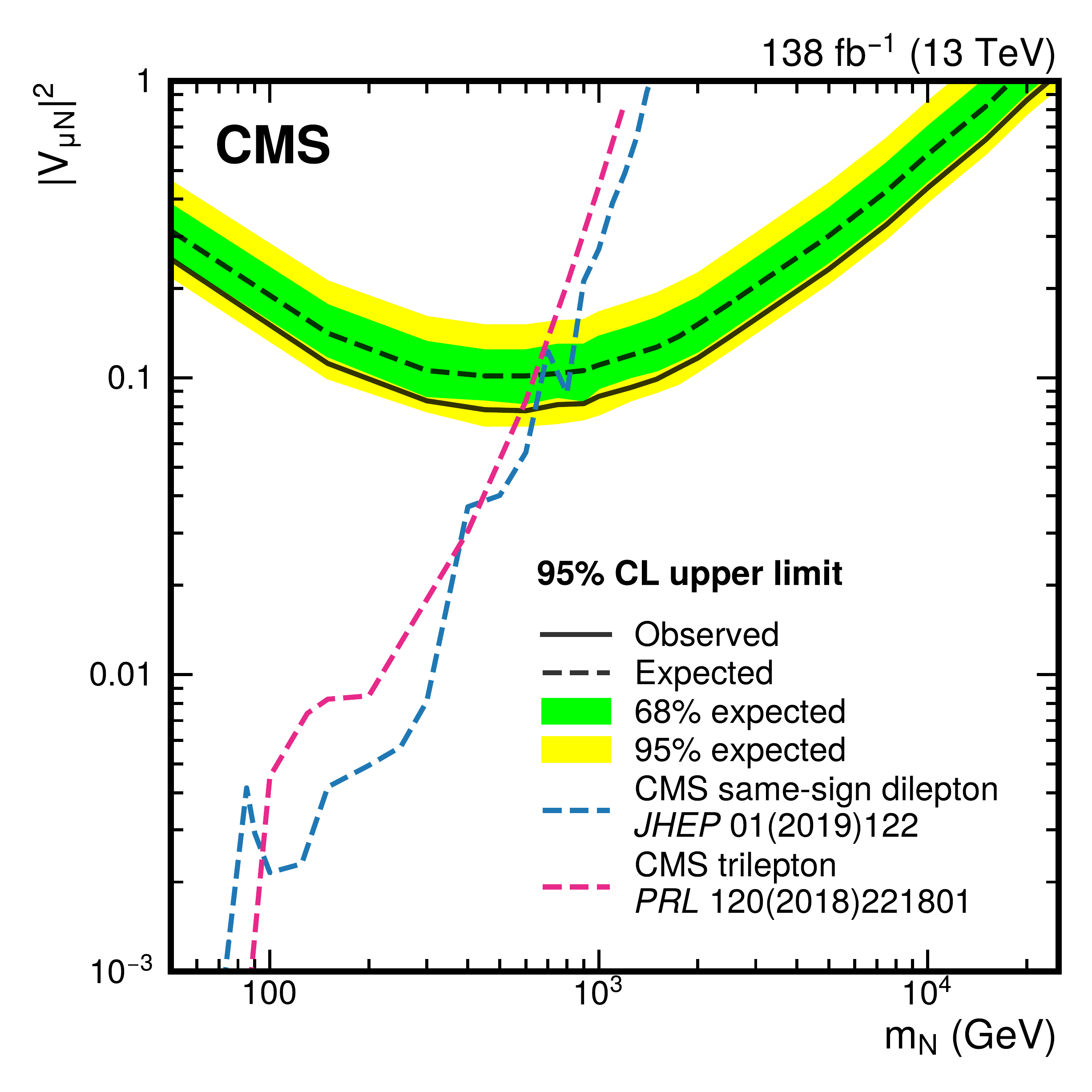}
    \caption{Upper limits on the heavy neutrino mixing element $\abs{ \Vmn }^{2}$ at the 95\% C.L. as a function of the heavy neutrino mass $\mN$. The black dashed curve shows the expected upper limit, with the 1 and 2 standard deviation bands are indicated in lime green and light yellow, respectively. The solid black curve is the observed upper limit. The red dashed curve indicates observed upper limits from Ref.~\cite{Sirunyan:2018mtv}, while the blue dashed curve shows the observed upper limits from Ref.~\cite{Sirunyan:2018xiv}. The result from Ref.~\cite{Aad:2019kiz} is not shown since it focuses on low $\mN$. Starting from $\mN$ around 650\GeV, the analysis presented in this Letter improves upon the upper limits from those references.}
    \label{fig:upper_limits}
\end{figure}

The data are found to be consistent with the background expectation. Using the relationship between the cross section and the mixing elements for the \hmn analysis, upper limits at the 95\% \CL are derived on $\abs{ \Vmn }^{2}$ as shown in Fig.~\ref{fig:upper_limits}. These results surpass those obtained in previous searches by the ATLAS and CMS Collaborations~\cite{Sirunyan:2018mtv,Sirunyan:2018xiv,Aad:2019kiz} for $\mN \gtrsim 650\GeV$, and set the first direct limits for $\mN > 2 \TeV$. Previous upper limits on the mixing elements from the CMS Collaboration~\cite{Sirunyan:2018mtv,Sirunyan:2018xiv} are plotted in Fig.~\ref{fig:upper_limits} for comparison. These limits on $\abs{ \Vmn }^{2}$ approach unity at around 1\TeV. As the mixing element describes the oscillation probability between the SM neutrinos and the heavy Majorana neutrino, results with $\abs{ \Vmn }^{2} > 1$ are not physically meaningful.

Equation~(\ref{eqn:Majorana_N_mass}) is used to convert the limit on $\abs{\mathrm{C}_5^{\ell\ell'}/\Lambda }^{2}$ extracted from the Weinberg operator analysis into a limit on $\abs{m_{\PGm\PGm}}$. The observed (expected) 95\% \CL upper limit on $\abs{m_{\PGm\PGm}}$ is found to be $10.8 ~ (12.8)\GeV$. This is the first upper limit from a collider experiment, and it improves upon a previous limit set in Ref.~\cite{Fuks:2020zbm} using the results from the NA62 experiment~\cite{NA62:2019eax}.
\ifthenelse{\boolean{cms@external}}{\clearpage}{}

In summary, this Letter presents the first search for Majorana neutrinos at several \TeVns{} and a first probe of the Weinberg operator at the LHC. These achievements were made possible by considering for the first time vector boson fusion processes resulting in a same-sign dimuon final state. The results are consistent with the predictions from the standard model.
For heavy Majorana neutrinos, upper limits on the mixing element $\abs{ \Vmn }^{2}$ are set for the mass range $50\GeV < \mN < 25\TeV$ and the best sensitivity is reached for $\mN \gtrsim 650\GeV$. The phase space explored exceeds the center-of-mass energy of the LHC, improving previous limits for direct production of Majorana neutrinos. The highest mass for which $\abs{\Vmn}^{2}=1$ is excluded is around 23\TeV. The observed (expected) 95\% confidence level upper limit on the effective $\PGm\PGm$ Majorana mass associated with the Weinberg operator is 10.8 (12.8)\GeV, exceeding the current best limit from high intensity Kaon experiments.

\begin{acknowledgments}
We congratulate our colleagues in the CERN accelerator departments for the excellent performance of the LHC and thank the technical and administrative staffs at CERN and at other CMS institutes for their contributions to the success of the CMS effort. In addition, we gratefully acknowledge the computing centers and personnel of the Worldwide LHC Computing Grid and other centers for delivering so effectively the computing infrastructure essential to our analyses. Finally, we acknowledge the enduring support for the construction and operation of the LHC, the CMS detector, and the supporting computing infrastructure provided by the following funding agencies: BMBWF and FWF (Austria); FNRS and FWO (Belgium); CNPq, CAPES, FAPERJ, FAPERGS, and FAPESP (Brazil); MES and BNSF (Bulgaria); CERN; CAS, MoST, and NSFC (China); MINCIENCIAS (Colombia); MSES and CSF (Croatia); RIF (Cyprus); SENESCYT (Ecuador); MoER, ERC PUT and ERDF (Estonia); Academy of Finland, MEC, and HIP (Finland); CEA and CNRS/IN2P3 (France); BMBF, DFG, and HGF (Germany); GSRI (Greece); NKFIH (Hungary); DAE and DST (India); IPM (Iran); SFI (Ireland); INFN (Italy); MSIP and NRF (Republic of Korea); MES (Latvia); LAS (Lithuania); MOE and UM (Malaysia); BUAP, CINVESTAV, CONACYT, LNS, SEP, and UASLP-FAI (Mexico); MOS (Montenegro); MBIE (New Zealand); PAEC (Pakistan); MES and NSC (Poland); FCT (Portugal); J MESTD (Serbia); MCIN/AEI and PCTI (Spain); MOSTR (Sri Lanka); Swiss Funding Agencies (Switzerland); MST (Taipei); MHESI and NSTDA (Thailand); TUBITAK and TENMAK (Turkey); NASU (Ukraine); STFC (United Kingdom); DOE and NSF (USA).
\end{acknowledgments}

\bibliography{auto_generated}

\cleardoublepage \appendix\section{The CMS Collaboration \label{app:collab}}\begin{sloppypar}\hyphenpenalty=5000\widowpenalty=500\clubpenalty=5000\input{EXO-21-003-public-authorlist.tex}\end{sloppypar}
\end{document}

%% file: EXO-21-003-public-authorlist.tex
\cmsinstitute{Yerevan Physics Institute, Yerevan, Armenia}
{\tolerance=6000
A.~Tumasyan\cmsAuthorMark{1}\cmsorcid{0009-0000-0684-6742}
\par}
\cmsinstitute{Institut f\"{u}r Hochenergiephysik, Vienna, Austria}
{\tolerance=6000
W.~Adam\cmsorcid{0000-0001-9099-4341}, J.W.~Andrejkovic, T.~Bergauer\cmsorcid{0000-0002-5786-0293}, S.~Chatterjee\cmsorcid{0000-0003-2660-0349}, K.~Damanakis\cmsorcid{0000-0001-5389-2872}, M.~Dragicevic\cmsorcid{0000-0003-1967-6783}, A.~Escalante~Del~Valle\cmsorcid{0000-0002-9702-6359}, P.S.~Hussain\cmsorcid{0000-0002-4825-5278}, M.~Jeitler\cmsAuthorMark{2}\cmsorcid{0000-0002-5141-9560}, N.~Krammer\cmsorcid{0000-0002-0548-0985}, L.~Lechner\cmsorcid{0000-0002-3065-1141}, D.~Liko\cmsorcid{0000-0002-3380-473X}, I.~Mikulec\cmsorcid{0000-0003-0385-2746}, P.~Paulitsch, F.M.~Pitters, J.~Schieck\cmsAuthorMark{2}\cmsorcid{0000-0002-1058-8093}, R.~Sch\"{o}fbeck\cmsorcid{0000-0002-2332-8784}, D.~Schwarz\cmsorcid{0000-0002-3821-7331}, S.~Templ\cmsorcid{0000-0003-3137-5692}, W.~Waltenberger\cmsorcid{0000-0002-6215-7228}, C.-E.~Wulz\cmsAuthorMark{2}\cmsorcid{0000-0001-9226-5812}
% ...
\par}
\vskip\cmsinstskip
\dag:~Deceased\\
$^{1}$Also at Yerevan State University, Yerevan, Armenia\\
$^{2}$Also at TU Wien, Vienna, Austria\\
% ...
